\documentclass[aps,amsfonts,pra,twocolumn,showpacs]{revtex4-1}

\usepackage{subfigure}
\usepackage{textcomp}
\usepackage{graphicx}
\usepackage{amssymb}
\usepackage{amsmath}
\usepackage{bm}
\usepackage{amsmath, amsthm, amssymb}
\usepackage{dsfont}
\usepackage[colorlinks=true,linkcolor=blue,urlcolor=blue,citecolor=blue]{hyperref}
\usepackage{multirow}
\usepackage{url}

\def\beq{\begin{equation}}
\def\eeq{\end{equation}}
\def\bea{\begin{eqnarray}}
\def\eea{\end{eqnarray}}

\begin{document}

\title{Facilitated quantum cellular automata as simple models with nonthermal eigenstates and dynamics}

\author{Sarang Gopalakrishnan and Bahti Zakirov}
\affiliation{Department of Engineering Science and Physics, CUNY College of Staten Island, Staten Island, NY 10314}
\affiliation{Physics Program and Initiative for the Theoretical Sciences, The Graduate Center, CUNY, New York, NY 10016}

\begin{abstract}
We introduce and describe a class of simple facilitated quantum spin models in which the dynamics is due to the repeated application of unitary gates. The gates are applied periodically in time, so their combined action constitutes a Floquet unitary. The dynamics of the models we discuss can be classically simulated, and their eigenstates classically constructed (although they are highly entangled). We consider a variety of models in both one and two dimensions, involving Clifford gates and Toffoli gates. For some of these models, we explicitly construct conserved densities; thus these models are ``integrable.'' The other models do not seem to be integrable; yet, for some system sizes and boundary conditions, their eigenstate entanglement is strongly subthermal. Some of the models have exponentially many eigenstates in which one or more sites are ``disentangled'' from the rest of the system, as a consequence of reflection symmetry.  
\end{abstract}

\maketitle

\section{Introduction}

The nonequilibrium dynamics of isolated, interacting quantum systems is a central theme in contemporary many-body physics~\cite{cazalilla2010focus, polk_RMP, nhreview}. 
Recently this theme has been explored from various perspectives, both experimental (particularly in ultracold atomic gases~\cite{langen2013local, kaufman2016quantum}) and theoretical (concerning the nature of quantum chaos~\cite{maldacena2016bound}, thermalization~\cite{nhreview}, and glassiness). 
Generic quantum systems (except for the many-body localized phase~\cite{BAA, nhreview}) are expected to approach local thermal equilibrium starting from essentially any initial state~\cite{deutsch_eth, srednicki_eth, ETH_outliers, dymarsky2017}. The locally equilibrated state is highly entangled~\cite{deutsch_eth}; the growth of entanglement, and the scrambling of local information, have been explored in various settings~\cite{tsunami, kimhuse, mezei2017, nrvh, curtvonK, xu2018accessing}. 
However, there are also less generic types of behavior, such as integrable dynamics: integrable systems are characterized by an infinite number of quasilocal conserved charges (i.e., quasilocal operators $O_i$ such that $\sum_i O_i$ commutes with the time evolution operator~\cite{grady, ilievski2016}). 
In principle, a system might fail to thermalize for reasons distinct from localization or integrability; this possibility has been less studied (but see the proposal in Ref.~\cite{qdl}).


In the present work we introduce a class of theoretically tractable models that have rich nonequilibrium dynamics, as well as highly nonthermal eigenstate entanglement patterns. Some of these models are integrable; others do not appear to be. Our models consist of time-periodic, translation-invariant quantum circuits, and are analogues of certain kinetically constrained models of glasses~\cite{ritort2003, garrahan2017aspects}. All but one of the models considered here involve Clifford gates~\cite{gottesman}. Random Clifford circuits were recently investigated~\cite{nrvh, clclifford}, but our aim here is different: we consider translationally invariant circuits in which the gates are applied in a time-periodic manner. Thus the dynamics is described by a translation-invariant Floquet unitary operator, allowing us to define and study properties such as many-body eigenstate entanglement. Our models are examples of Clifford quantum cellular automata~\cite{schumacher2004, schlingemann2008, gutschow2010, werner2010fractal}; some general properties of the dynamics of such automata are known, but eigenstate thermalization, operator spreading, and related questions have not been addressed for this class of models. 

We consider five models. The first four consist entirely of Clifford gates: specifically, periodically applied patterns of controlled NOT (CNOT) gates. Two of these models live in one dimension; the two others, on square and Kagome lattices respectively. The fifth model involves both CNOT and Toffoli gates~\cite{nielsen_chuang}. All models share the very nongeneric property that unitary evolution maps product states in the computational basis to other product states, not superpositions. In some sense, therefore, these models are ``classical''; nevertheless, the eigenstates are highly entangled, and we find all three classes of behavior that were predicted to exist in Clifford cellular automata~\cite{gutschow2010}. 
Two of our models are integrable; we construct their conserved charges. The other three (including the CNOT-Toffoli model) do not seem to be integrable; yet the dynamics of all these models deviates from our expectations regarding generic chaotic quantum dynamics. 
Even in the nonintegrable models, we find properties such as self-similar operator growth inside the light-cone, and a finite entropy of many-body eigenstates with one or more ``disentangled'' sites, which strongly violate the eigenstate thermalization hypothesis~\cite{deutsch_eth, srednicki_eth, Rigol:2008kq}.

The rest of this paper is organized as follows. In Sec.~\ref{background} we briefly review the notions of kinetically constrained models, Clifford gates and automata, and eigenstate thermalization. In Sec.~\ref{summary} we outline our approach to constructing eigenstates and bounding their entanglement, and introduce the various models that we then analyze in Secs.~\ref{east}-\ref{toffsec}. Sec.~\ref{edresults} presents some results obtained from the exact diagonalization of small systems (which is more convenient for addressing some specific observables). Finally, Sec.~\ref{discussion} closes with a summary of our results and their broader relevance.

\section{Background}\label{background}


\subsection{Kinetically constrained models}

Facilitated, or kinetically constrained, models (KCMs) are a class of models that were introduced to explain the glass transition~\cite{ritort2003}. The simplest classical KCMs consist of a lattice of classical two-state systems (``spins''), with a dynamical rule that each spin can flip if its neighbors satisfy some condition. Such models include the East model, in which a spin can flip only if its left neighbor is up, and the Frederickson-Andersen model, in which a spin can flip if either of its neighbors is up (but not if both are down)~\cite{ritort2003}. In these models the dynamics of a site depends strongly on the state of its immediate environment, and is thus spatially heterogeneous for a random initial state. Hamiltonian quantum versions of these models can be constructed with the use of projection operators: e.g., for the East model, a Hamiltonian such as $H = \sum_i (1 - \sigma^z_i) \sigma^x_{i+1}$ would give rise to similar dynamics. The quantum East model has been studied from this perspective in Ref.~\cite{hickey2016}.

\subsection{Clifford gates and automata}

Clifford gates~\cite{gottesman} are a set of quantum gates (i.e., local unitary operators) with the following property: in the Heisenberg picture, a Clifford gate maps each string of Pauli operators to a \emph{unique} string of Pauli operators (by contrast, generic gates map a Pauli string to a \emph{superposition} of Pauli strings). For time evolution consisting purely of Clifford gates, the dynamics of Pauli operators (and therefore of initial product states in the computational basis) can efficiently be simulated on a classical computer. For circuits consisting of Clifford gates, one can also define the notion of ``Clifford cellular automata,'' i.e., block cellular automata that act on the \emph{operators} rather than the states, updating them by applying Clifford gates~\cite{schumacher2004, schlingemann2008}. Such automata have been discussed and classified from a mathematical perspective~\cite{gutschow2010, werner2010fractal}; this work anticipates some of our results, in a more general setting, but does not address the aspects of thermalization with which we are concerned here.

\subsection{Thermalization and operator spreading}

It is believed that a generic isolated quantum system, started far from equilibrium, will rapidly approach local equilibrium under its unitary dynamics. This belief is formalized as the eigenstate thermalization hypothesis (ETH)~\cite{deutsch_eth, srednicki_eth}. Although ETH was originally formulated for Hamiltonian systems, it naturally extends to any unitary dynamics in which the unitary is periodic in time (``Floquet'' unitaries), and it is this broader version that concerns us here~\cite{ETH_outliers}. ETH states that local observables in a many-body eigenstate take the same values as they would in a maximum-entropy density matrix subject to the same conservation laws and global constraints. (For a Hamiltonian system, energy is conserved; thus local observables take the maximum-entropy value consistent with the global energy density, i.e., the thermal value.) An implication of ETH is that the observed thermal entropy of a small subsystem is entirely due to its entanglement with the rest of the system; thus the entanglement entropy and thermal entropy match. However, the matching of entropies does not \emph{imply} ETH: general integrable systems, and even free-fermion systems, can exhibit entanglement volume laws. There are both weak and strong versions of ETH~\cite{garrison_grover, dymarsky2017, dymarsky2018}: the weaker forms posit only that local operators behave thermally, whereas the strongest forms posit that even subsystems that are a finite fraction of the full system size are described by thermal reduced density matrices. Much of the numerical evidence~\cite{ETH_outliers, garrison_grover, dymarsky2018} is consistent with these stronger forms of ETH.


Thermalization is associated with chaotic dynamics, which in turn is captured by the out-of-time-ordered correlator (OTOC): $\langle [A_i(t), B_j(0)]^2 \rangle$ for general local operators $A, B$ at lattice sites $i, j$~\cite{larkin_otoc, maldacena2016bound}. The OTOC is small when two degrees of freedom are causally disconnected, and grows when they come into causal contact, i.e., when the Heisenberg operator $A_i(t)$ has ``spread'' under the time dynamics to reach the point $j$. 
%
In generic thermal systems, the OTOC is uniformly large inside the causal light-cone of the operator~\cite{tsunami, mezei2017, nrvh} (the notion of light-cones is to be understood in terms of Lieb-Robinson bounds~\cite{lieb_robinson}) and small outside it. 

\section{Main ideas of this work} \label{summary}


The present work implements quantum KCMs as reversible cellular automata, in which each site is updated at every step by local unitary gates that map product states in the computational basis to other product states. 
We explore operator dynamics and eigenstate thermalization in these models. Because we are considering Floquet systems, energy is not conserved, and energetics does not constrain particle motion; rather, the only constraints on the dynamics are the kinetic constraints. 

\subsection{Constructing eigenstates}

The circuits we will discuss here are built up of controlled NOT (CNOT) gates. Thus our unitaries map computational-basis product states to product states, e.g., $|\psi(0) \rightarrow \psi(1) \rightarrow \ldots |\psi(n)\rangle \rightarrow |\psi(0) \rightarrow \ldots$. By following the trajectory of each product state through configuration space, we can construct Floquet eigenstates as $\sum_{k \leq n} e^{i \epsilon_n n} |\psi(n) \rangle$. (This construction extends to general Clifford circuits, since time evolution maps each product state to a unique Pauli operator-string applied to the computational vacuum state $|000\ldots 0\rangle$.)
%
Each configuration is evolved until it returns to its initial state; the size of the configuration-space orbit will be an important property for us, because of the following useful bound. 

\subsection{Participation ratios and entanglement}\label{lemma}

The inverse participation ratio (IPR) is an important quantity in the theory of localization; the configuration-space form we use here was introduced in Ref.~\cite{BAA} to study many-body localization. This quantity, which we denote $\mathcal{I}$, is defined as follows, for a normalized state $|\psi\rangle$:

\beq\label{ipr}
\mathcal{I}[\psi] = \sum_c |\langle c | \psi \rangle|^4,
\eeq
where $c$ labels the configurations of the system in the computational basis (which consists of product states in the $z$ basis). The inverse of $\mathcal{I}$, the ``participation ratio,'' counts the number of states accessible to the system, and (for our models) corresponds to the length of its classical orbit. The configuration-space IPR can be seen to bound the entanglement entropy (for real-space entanglement cuts), by the following logic. Suppose the state is spread out over $N$ configurations, and can be written as 

\beq\label{decomp}
|\psi\rangle = \sum_{i = 1}^N a_i |\psi_A\rangle \otimes |\psi_B\rangle,
\eeq
where $|\psi_{A,B}\rangle$ live on either side of the entanglement cut. The maximum possible entanglement occurs when all the $|\psi_A\rangle$ and all the $|\psi_B\rangle$ are mutually distinguishable; in this case the computational basis is the Schmidt basis~\cite{nielsen_chuang}. In this case the second Renyi entropy~\cite{Renyi} is given by

\beq
S_2[|\psi\rangle] = - \ln(|a_i|^4) = -\ln(\mathcal{I}[|\psi\rangle]).
\eeq 
Typically, there will be some repeated $|\psi_A\rangle$ and/or $|\psi_B\rangle$ in the decomposition~\eqref{decomp}. Repeats invariably decrease entanglement, so the following inequality generally holds for the second Renyi entropy:

\beq\label{Bound}
S_2[|\psi\rangle] \leq -\ln(\mathcal{I}[|\psi\rangle]).
\eeq
In general one should distinguish between the Renyi and Von Neumann entanglement entropies; however, these behave similarly for Clifford systems~\cite{nrvh}, and we shall treat them as interchangeable. For most systems the bound~\eqref{Bound} is quite loose, even for properties such as the half-system entanglement. However, in the present context it is simpler to count distinct configurations than to compute entanglement, and we shall find that the bound is in some cases tight enough to rule out volume-law entanglement for regions that occupy a finite fraction of the system.

\subsection{Models}

In the bulk of this paper we discuss the four models introduced below. Unless otherwise specified, we focus on bipartite systems subject to periodic boundary conditions.

\begin{figure}[tb]
\begin{center}
\includegraphics[width = 0.45\textwidth]{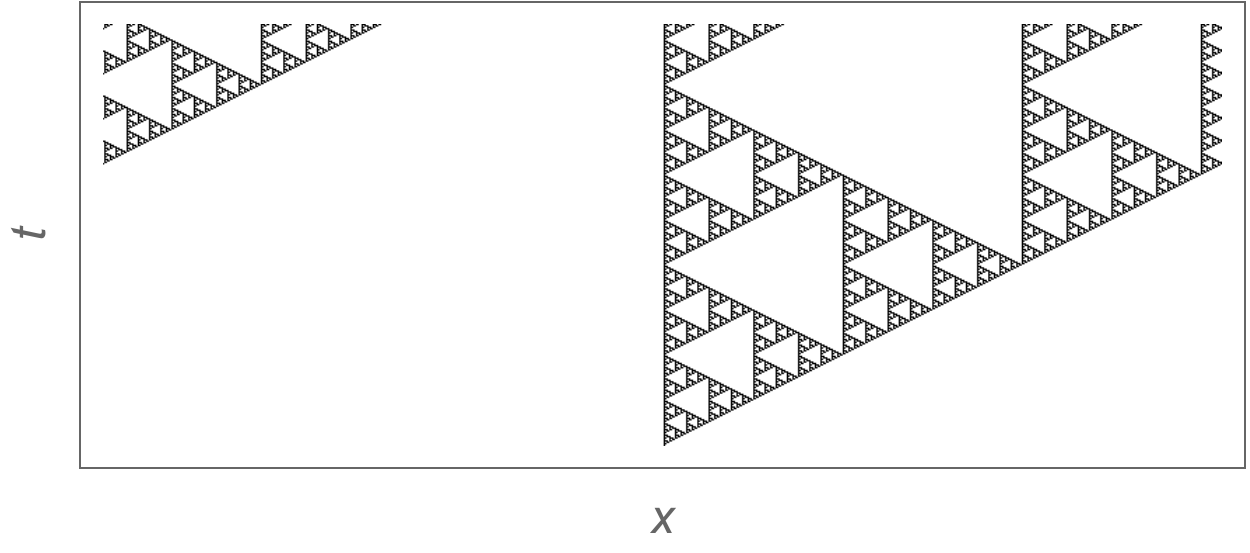}
\caption{Growth of the Pauli $X$ (i.e., $\sigma_x$) operator with time in the Clifford-East model~\eqref{model1}. }
\label{eastOG}
\end{center}
\end{figure}

\begin{enumerate}

\item \emph{Clifford-East Model}.---In this model one applies a controlled NOT (CNOT) gate from every spin to its rightward neighbor (i.e., one flips the neighbor if the control spin is up). Operators spread in self-similar patterns resembling Sierpinski triangles; moreover, operator spreading is highly anisotropic (Fig.~\ref{eastOG}). The $\hat{X}$ operator (i.e., the on-site Pauli matrix $\sigma_x$) spreads only to the right, and the $\hat{Z}$ operator spreads only to the left. All operators spread, and there are no local conserved quantities, so this model is not conventionally integrable. 
However, eigenstate entanglement is strongly dependent on system size. 
The size-dependence obeys the following pattern: for system sizes that are twice a prime number, generic eigenstates appear fully thermal, but for system sizes that are a power of two, the half-chain entanglement entropy of typical eigenstates scales at most as $\log L$, where $L$ is the system size. 

\item \emph{One-dimensional parity model}.---In this model one applies a CNOT gate to every spin from its rightward and then from its leftward neighbor. A spin is flipped if the left or the right neighbor is up, but not if both are up. This model is integrable: we explicitly construct an extensive set of conserved operators, consisting of the numbers of various left- and right-moving excitations. 
Unlike generic free fermion models, all excitations have the same velocity, so the recurrence time grows linearly with system size and the half-chain eigenstate entanglement entropy asymptotically grows as $\log L$.
 An interesting feature of this model is the presence of exponentially many disentangled sites: specifically, for an $L$-site chain (with $L$ even) there are $\sim L 2^{L/2}$ many-body eigenstates in which at least one site has zero entanglement entropy. This feature is due, not to the integrability of the model, but to its reflection symmetry, as we discuss below. 

\item \emph{Square-lattice parity model}.---The parity model generalizes naturally to bipartite lattices in higher dimensions. We first consider a square-lattice model. This model is not obviously integrable: all operators spread and there seem to be no conserved local operators. Yet operator dynamics remains nontrivial, and this model possesses exponentially many states with some disentangled spins. As in the Clifford-East model, eigenstate entanglement strongly depends on the prime factorization of the system size. 

\item \emph{Kagome-lattice parity model}.---Unlike the square-lattice parity model, the Kagome-lattice model (defined on a tripartite lattice, so that one applies CNOT gates first to the A sites, then the B and C sites, and so on) has extensively many conserved quantities. These are strictly spatially local, as in the many-body localized phase: they are products of $\hat{X}$ over individual hexagons of the lattice. 

\end{enumerate}

Finally we discuss a non-Clifford model (based on CNOT and Toffoli gates) for which, again, we can explicitly construct eigenstates and study their entanglement properties. This model appears to have chaotic operator spreading, but its eigenstate entanglement also violates the stronger forms of ETH: specifically the half-system entanglement entropy grows too slowly for strong ETH to hold.

\section{Clifford-East model}\label{east}

We first consider a one-dimensional model defined by the following Floquet unitary

\beq\label{model1}
U = \prod_{i\, \text{even}} \text{CNOT}(i \rightarrow i + 1)\prod_{j\, \text{odd}} \text{CNOT}(j \rightarrow j + 1).
\eeq
The rules for Heisenberg evolution of Pauli matrix strings under a CNOT gate are as follows (we only consider the two relevant sites where the gate acts): under $\text{CNOT}(1 \rightarrow 2)$, we have $I \otimes X \mapsto I \otimes X, X \otimes I \mapsto X \otimes X, I \otimes Z \mapsto Z \otimes Z, Z \otimes I \mapsto Z \otimes I$. 

\subsection{Operator spreading}

Under the unitary~\eqref{model1}, evidently, the operator $X$ spreads to the right and $Z$ to the left. The evolution of the operator $Y = i ZX$ is given by $U Y_i U^\dagger = i (U X_i U^\dagger) (U Z_i U^\dagger)$, and similarly for other products of Pauli operators. The operator $X$ spreads as a spacetime fractal, specifically a Sierpinski triangle (Fig.~\ref{eastOG}); it was argued in Ref.~\cite{gutschow2010, werner2010fractal} that this behavior is one of three generic possibilities for Clifford quantum cellular automata. 

For a general state, the timescale on which the commutator $[X_i(0), Z_{i+n}(t)]$ first grows large depends on the sign of $n$. If $n > 0$ the operator typically grows large after $n/2$ timesteps; for $n < 0$ the corresponding growth timescale is $|L - n|$ steps where $L$ is the system size. When the initial operator is $Z$ instead of $X$, an analogous result holds but with the signs flipped. The commutator $[Y_i(0), X_{i+n}(t)]$ behaves like $[Z_i(0), X_{i+n}(t)]$ whereas $[Y_i(0), Z_{i+n}(t)]$ behaves like $[X_i(0), Z_{i+n}(t)]$. The results above are for periodic boundary conditions; when the boundary conditions are open (or in the thermodynamic limit) ``wrong-side'' commutators never grow. As Fig.~\ref{eastOG} shows, even after the light-cone passes through a point, the commutator does not remain large; instead it grows and shrinks self-similarly inside the light-cone. 

It seems that there should be no local conserved densities in this model, as all local Pauli strings necessarily spread. Consider $X$ strings for simplicity. The left end of a local operator is fixed until the operator ``wraps around'' the system, so time evolution (over timescales of order unity) cannot translate the string. The right end of a local operator, on the other hand, generically grows, so time evolution cannot leave the string localized. Thus the only option is for the string to spread. 

\subsection{Many-body eigenstates}

We construct many-body eigenstates as discussed in Sec.~\ref{summary}: we follow the classical evolution of a state in the computational basis and sum over its ``orbit.'' This construction immediately yields the configuration-space IPR of an eigenstate. For small systems we have compared this procedure with explicit exact diagonalization, and found perfect agreement. As discussed in Sec.~\ref{lemma} the configuration-space IPR upper-bounds the entanglement entropy across any spatial cut: $S_{max} \equiv -\log(\mathcal{I}) \geq S_2$. %
In addition, we can use our knowledge of the eigenstates to efficiently compute their entanglement entropy. To make contact with our results on the IPR, we compute the second Renyi entropies of eigenstates derived from time-evolving random initial configurations. 

\begin{figure}[tb]
\begin{center}
\begin{minipage}{0.2\textwidth}
\includegraphics[width=\linewidth]{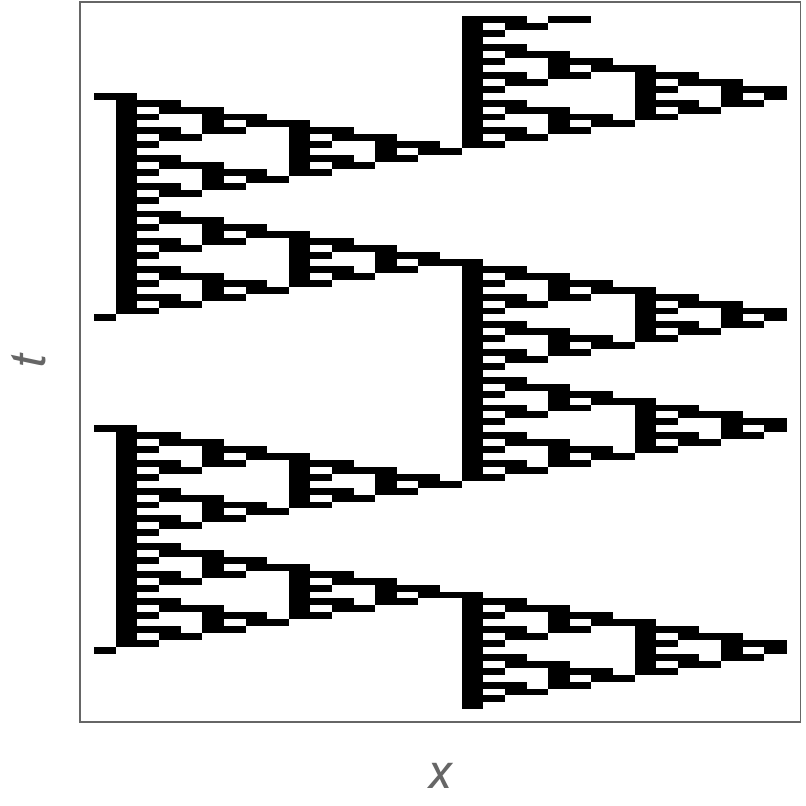}
\end{minipage}
\begin{minipage}{0.2\textwidth}
\includegraphics[width=\linewidth]{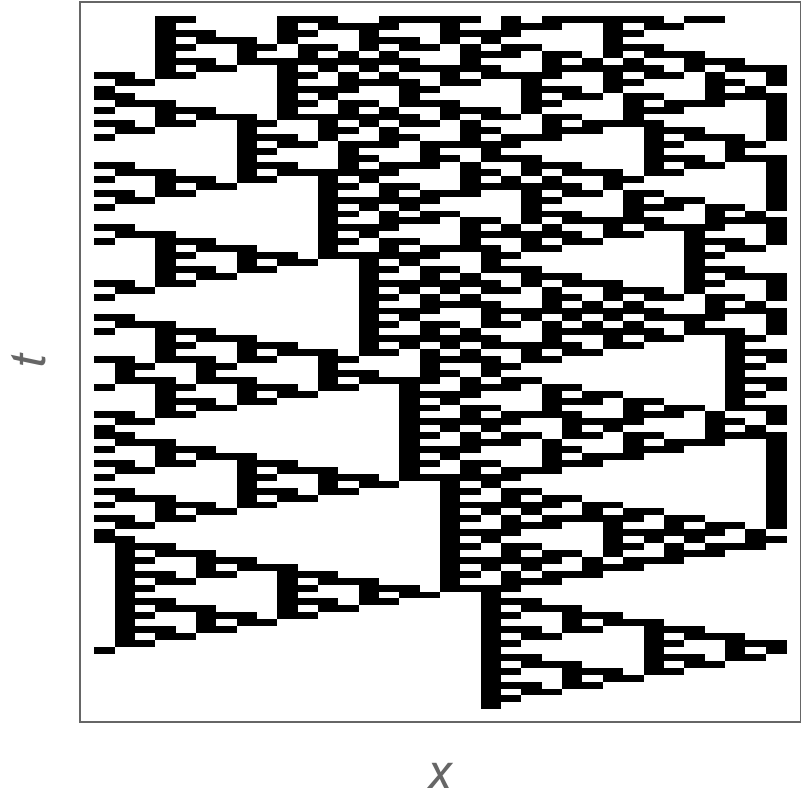}
\end{minipage}
\caption{Time evolution of the operator $X_i$ for system sizes $L = 32$ (left) and $L = 34$ (right). Note the very different behavior on timescales when information has wrapped around the system multiple times.}
\label{eastfig2}
\end{center}
\end{figure}

Our results for the IPR and Renyi entropy at various $L$ are shown in Fig.~\ref{eastfig2}. The randomly sampled Renyi entropies match exact diagonalization results for system sizes $L \leq 12$; in addition, the IPRs and Renyi entropies follow the same trends with system size. A striking feature is that for $L = 2^q$, $S_2$ grows at most \emph{logarithmically} with system size; more generally, for any length $L$, the relation $S_{max}(2L) = S_{max}(L) + \log 2$ appears to hold. As one can see from the time dynamics (Fig.~\ref{eastfig2}), this system-size dependence is essentially a commensurability effect: for the ``non-chaotic'' lengths, the system returns to its initial configuration after a few cycles, whereas for ``chaotic'' lengths it does not, and instead loops through a significant fraction of configuration space. 

\begin{figure}[tb]
\begin{center}
\includegraphics[width=0.4\textwidth]{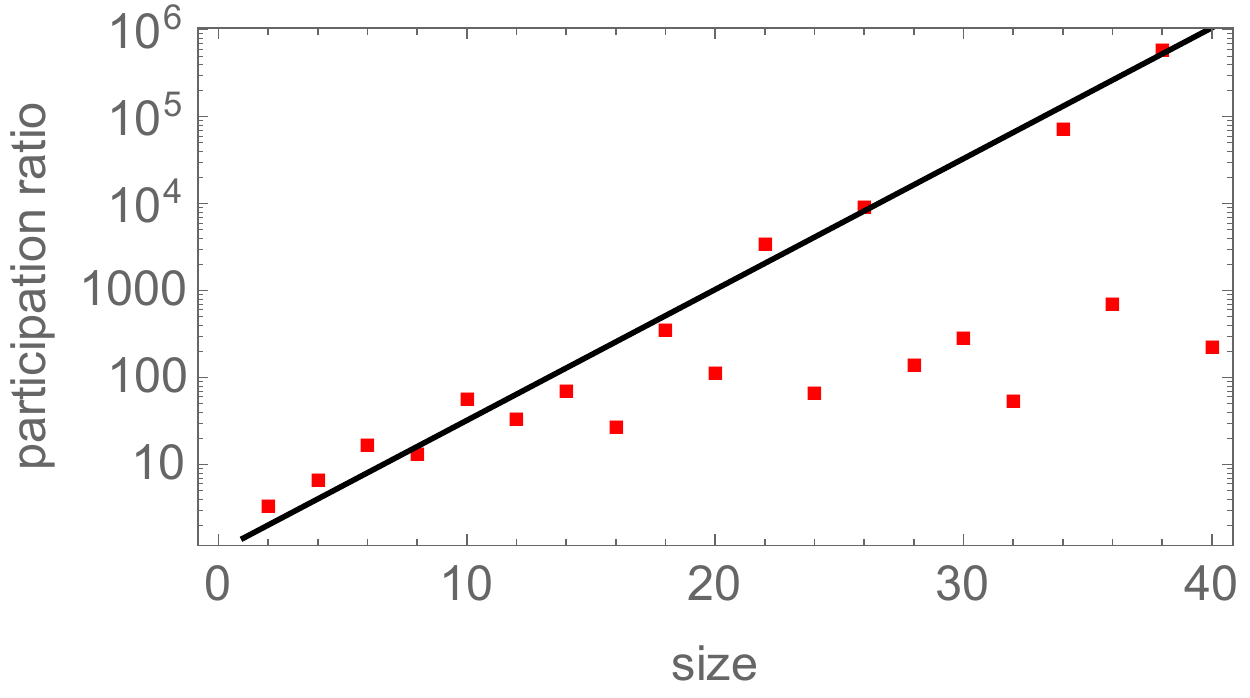}
\begin{minipage}{0.22\textwidth}
\includegraphics[width=\linewidth]{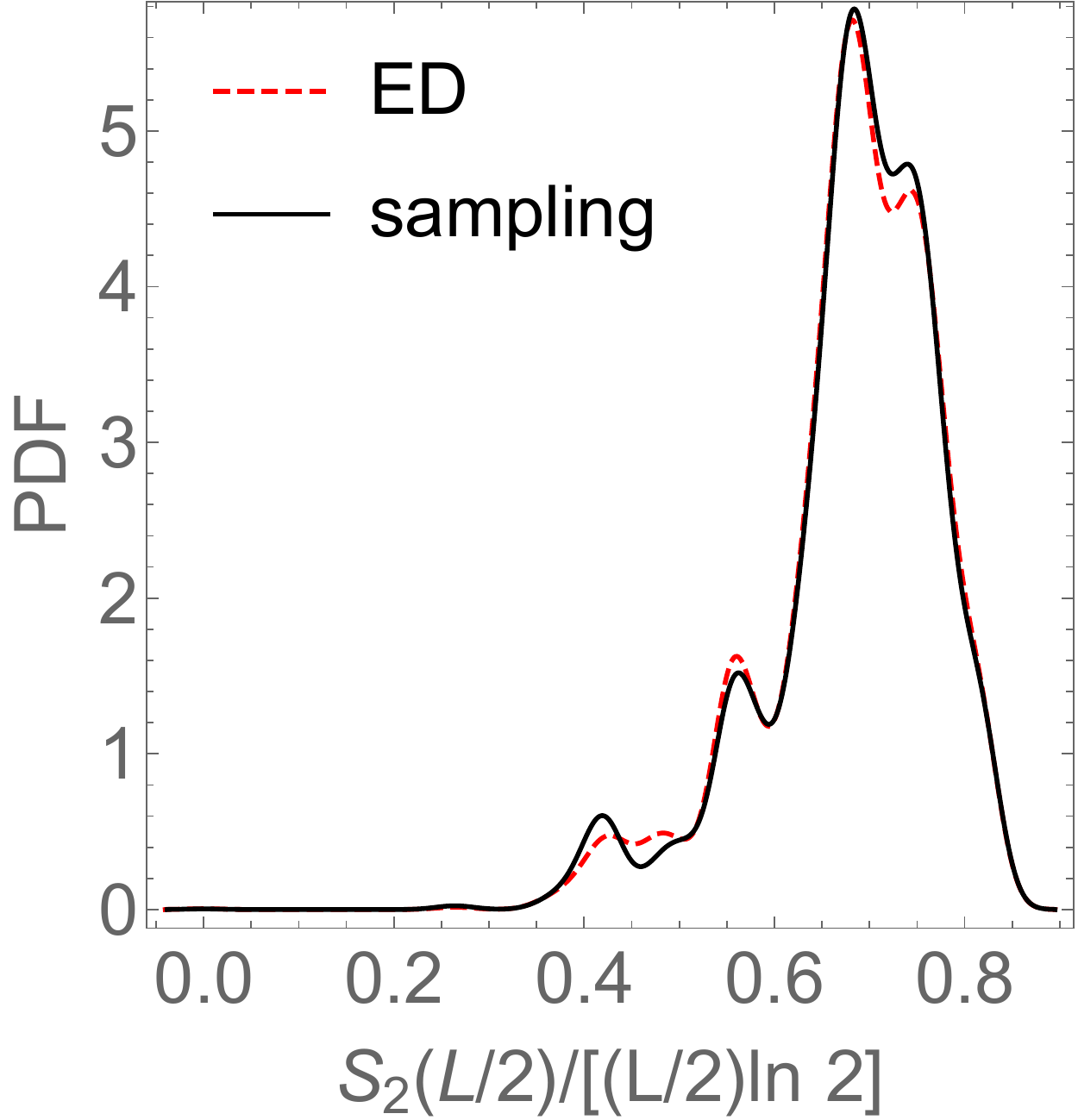}
\end{minipage}
\begin{minipage}{0.23\textwidth}
\includegraphics[width=\linewidth]{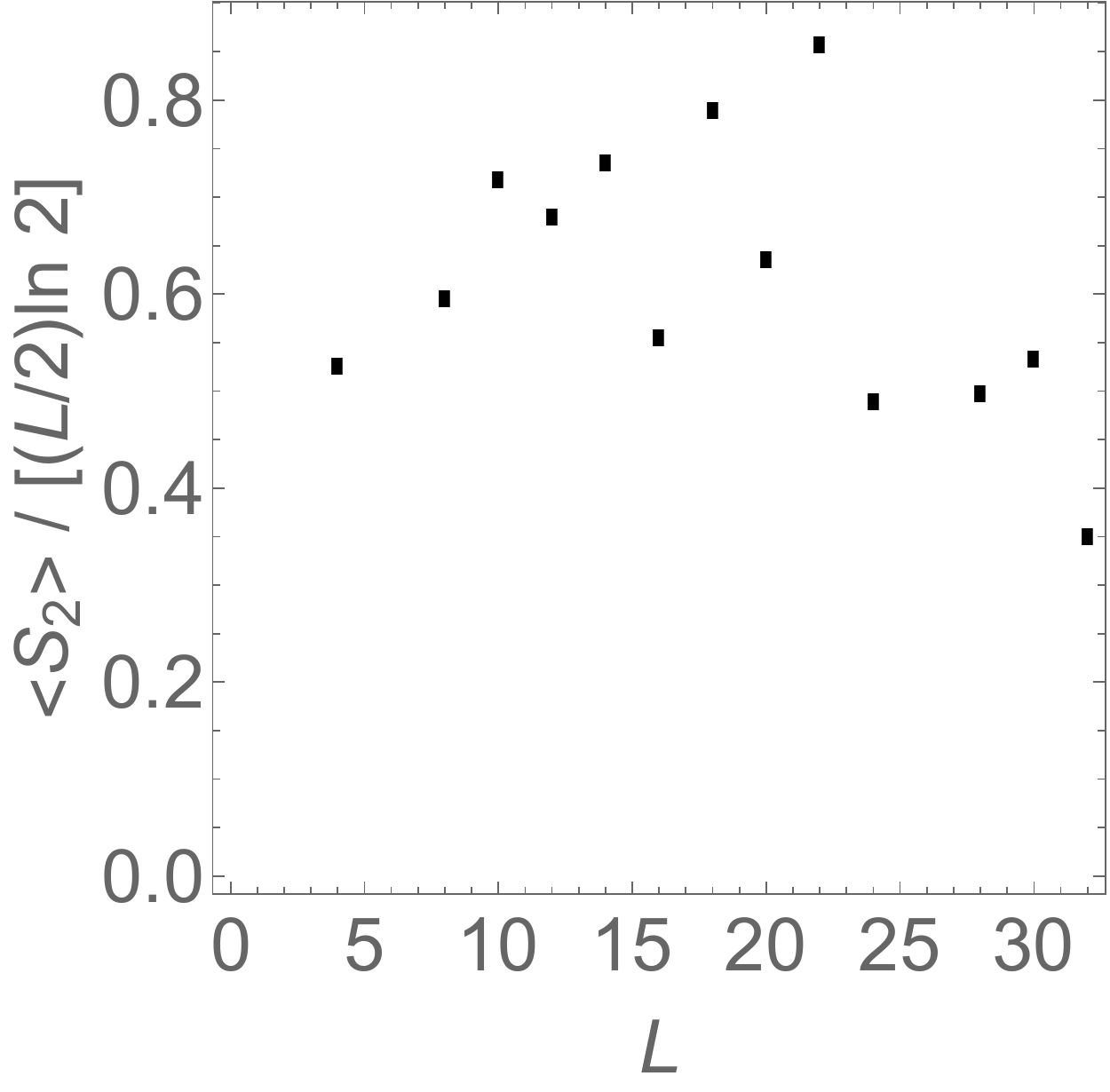}
\end{minipage}
\caption{(Upper panel) Configuration space IPRs for the Clifford-East model vs. system size (red dots). The black line denotes the smallest IPR consistent with maximal half-chain entanglement entropy. Left: comparison of histograms of the half-chain Renyi entropy $S_2(L/2)$ between full diagonalization of $L = 12$ systems and random classically constructed eigenstates. Right: $S_2(L/2)$ averaged over 100 states, normalized by its infinite-temperature value, vs. system size.}
\label{renyis}
\end{center}
\end{figure}

These observations apply to \emph{typical} initial states; there are, however, special initial states that have substantially less entanglement. The state with all spins down (which is a trivial eigenstate) is an example; so are ``blinker'' states in which (e.g.) all the $A$-sites are up and all the $B$-sites are down. We also find that the single-site entanglement entropy for regions much smaller than the half-chain grows as the volume of the region. Generally we expect the entanglement entropy in this model to have two behaviors: for regions much smaller than system size, we expect a volume law, which crosses over to logarithmic growth at a scale corresponding to $\log L$. This is consistent with the small-system diagonalization results in Sec.~\ref{edresults}.

\subsection{Generalizations}

We have also explored generalizations of this model in which CNOT gates are applied to a spin both from its nearest and next-nearest neighbor to the left: i.e., a spin is flipped if either its nearest neighbor or its next nearest neighbor (but not both) are up. This model behaves qualitatively similarly to the Clifford-East model, but since the lattice is not bipartite, $X$ and $Z$ no longer grow as mirror images of one another. Regardless, both operators exhibit the same qualitative fractal growth.

\section{One-dimensional parity model}\label{1dparity}

We now turn to the one-dimensional model defined by the following Floquet unitary:

\beq\label{model1}
U = \prod_{\pm, i\, \text{even}} \text{CNOT}(i \rightarrow i \pm 1)\prod_{\pm, j\, \text{odd}} \text{CNOT}(j \rightarrow j \pm 1).
\eeq
This model is a variant on one previously explored by Carr and coworkers~\cite{carrgroup}. In Ref.~\cite{gutschow2010} similar models are mapped to free fermions, and most likely this is possible in the present case also. In fact, the integrability of the present model is straightforward to see in the spin language, as described below. 

\subsection{Operator spreading}

Even-length strings of either $X$ or $Z$ simply translate by two steps under the unitary. Thus operators like $\sum_{i \, \text{even}} X_i X_{i+1}$ or $\sum_{j \, \text{odd}} Z_j Z_{j+1} Z_{j+2} Z_{j+3}$ commute with time evolution. In the cellular-automata language these operators are called ``gliders''~\cite{gutschow2010}. For our purposes we can equivalently regard them as conserved densities. There are two inequivalent classes of these conserved operators, depending on whether they are built out of $X$ or $Z$, and on whether they start at an even site or an odd site at the beginning of a cycle. $X$-strings starting at an odd site (or $Z$-strings starting at an even site) move to the right, whereas the other two types of strings move to the left. (Changing the origin of the Floquet cycle by half a period interchanges the two inequivalent $X$-type or $Z$-type strings). Odd-length strings are ``breathers,'' expanding and then contracting uniformly. 

\subsection{Eigenstate entanglement and disentangled sites}

Unlike generic free fermion models, in the present model all excitations move with the same speed. Thus, a state returns to itself after cycling through only $L/2$ configurations. The half-chain entanglement is thus bounded at $S_{max} = \frac{1}{2}\log L$ for all system sizes. Even local properties are strongly nonergodic: a striking example of this is the presence of ``disentangled'' sites in many eigenstates, i.e., sites that have no entanglement with the rest of the chain. The origin of disentangled sites can be understood as follows: take a configuration that is reflection-symmetric about some particular site. Under time evolution the two spins near that site will both be either spin-up or spin-down, and thus the site will never flip. Since this is true for any reflection-symmetric configuration, and there are $o(2^{L/2})$ configurations symmetric about a particular point, there are $o(L \times 2^{L/2})$ eigenstates in which at least one spin is disentangled. 




\section{Square-lattice parity model}\label{squaresec}

We now turn to a two-dimensional model, defined on a square lattice, with the same basic nature as the one-dimensional parity model. This unitary takes the form

\begin{figure}[tb]
\begin{center}
\begin{minipage}{0.23\textwidth}
\includegraphics[width =0.75 \linewidth]{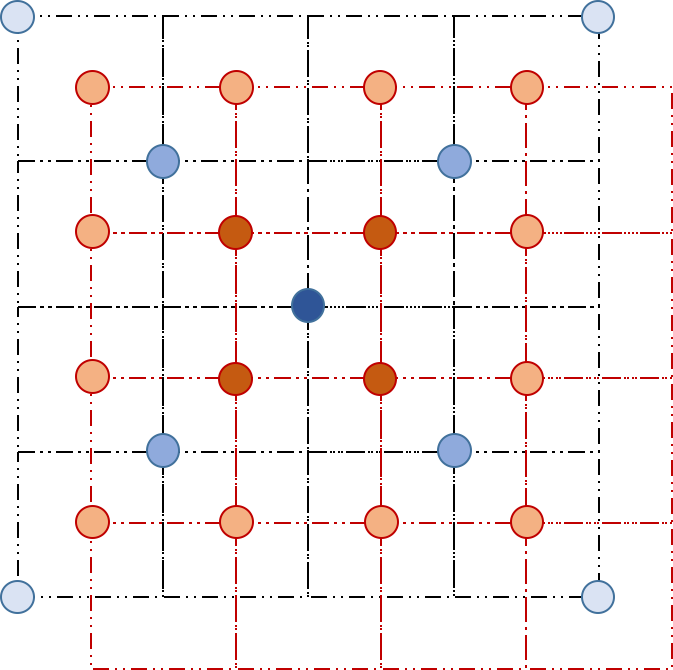}
\end{minipage}
\begin{minipage}{0.23\textwidth}
\includegraphics[width=0.8\linewidth]{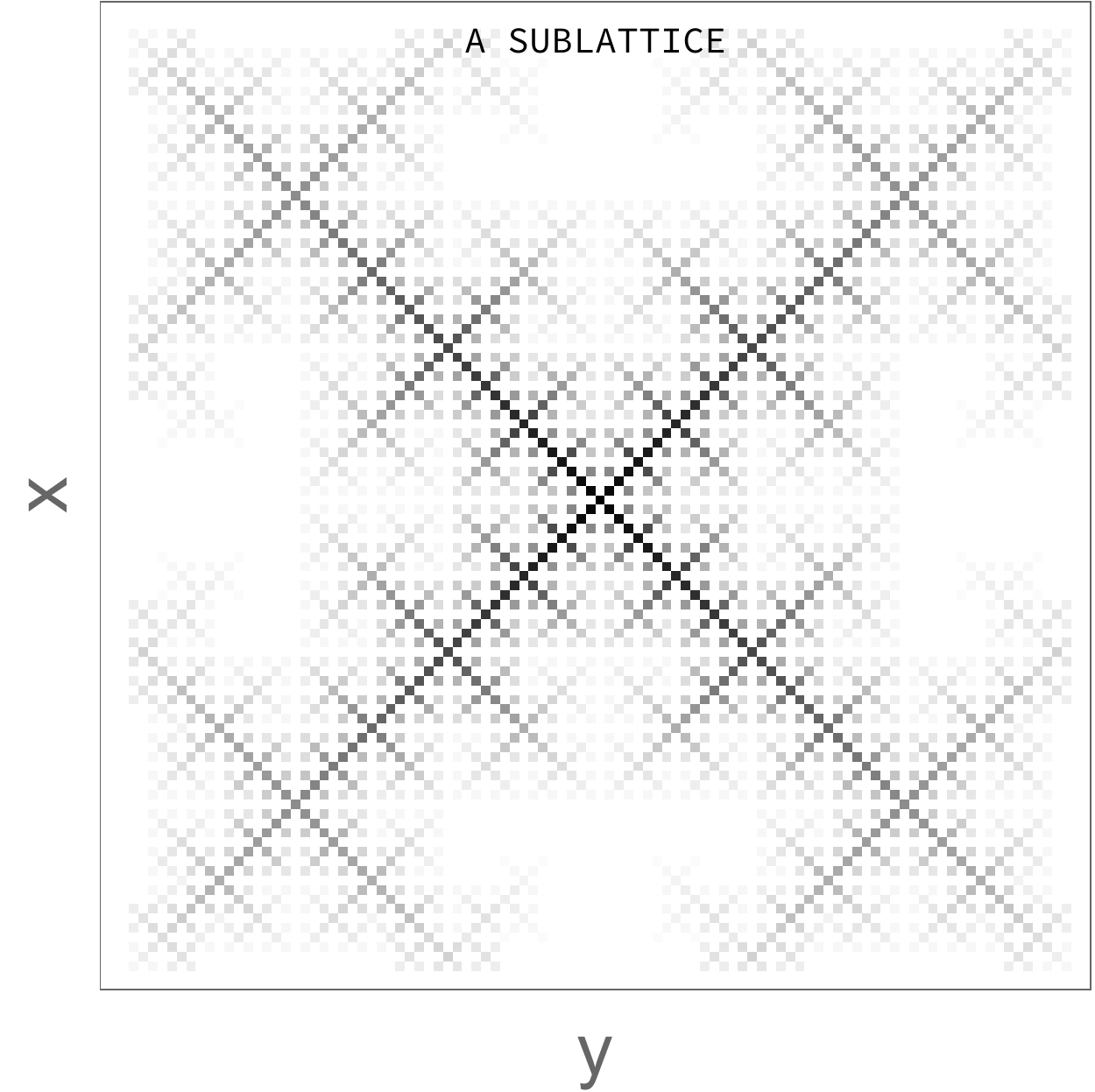}
\end{minipage}
\newline
\begin{minipage}{0.23\textwidth}
\includegraphics[width =0.8 \linewidth]{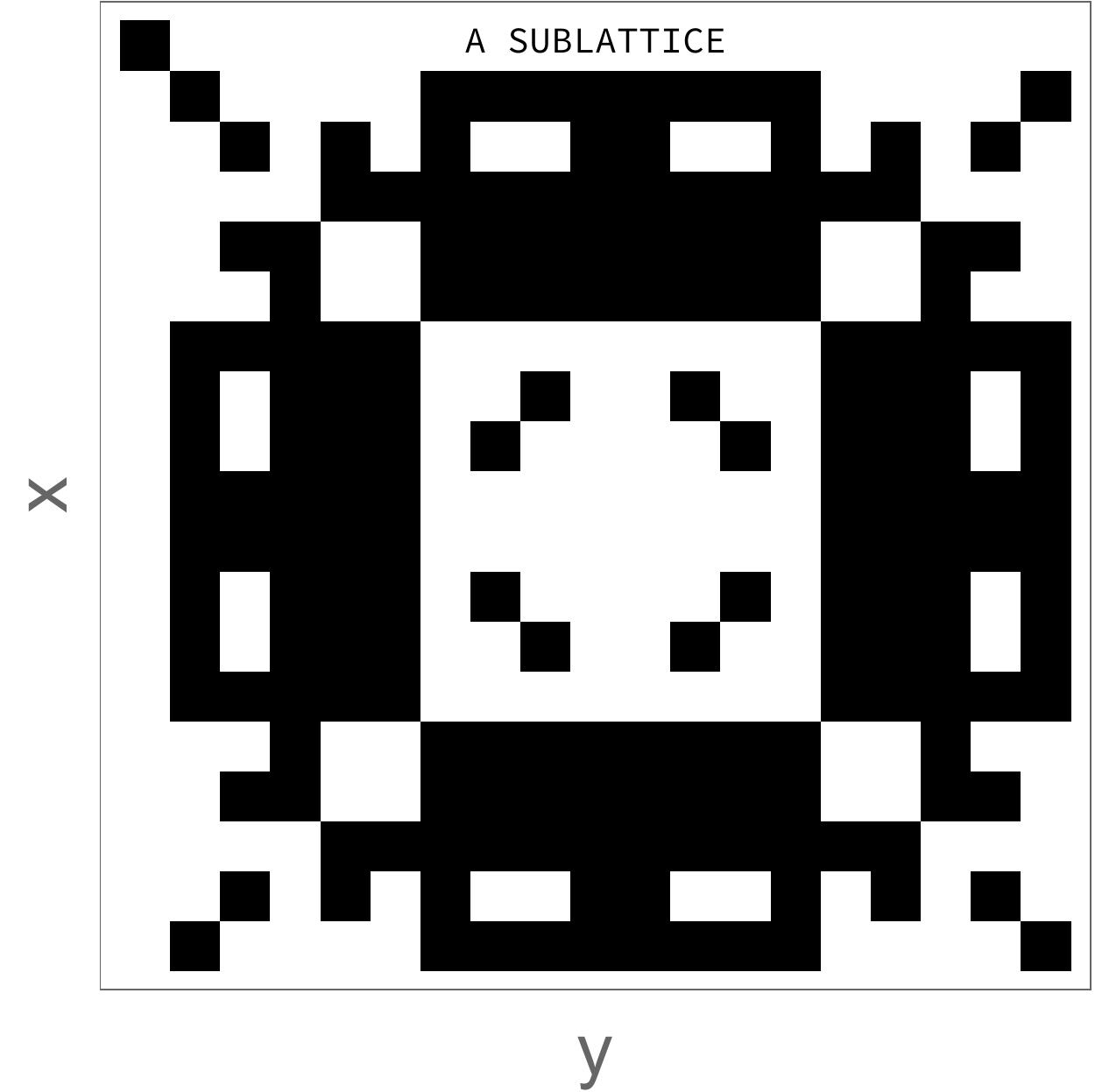}
\end{minipage}
\begin{minipage}{0.23\textwidth}
\includegraphics[width=0.9\linewidth]{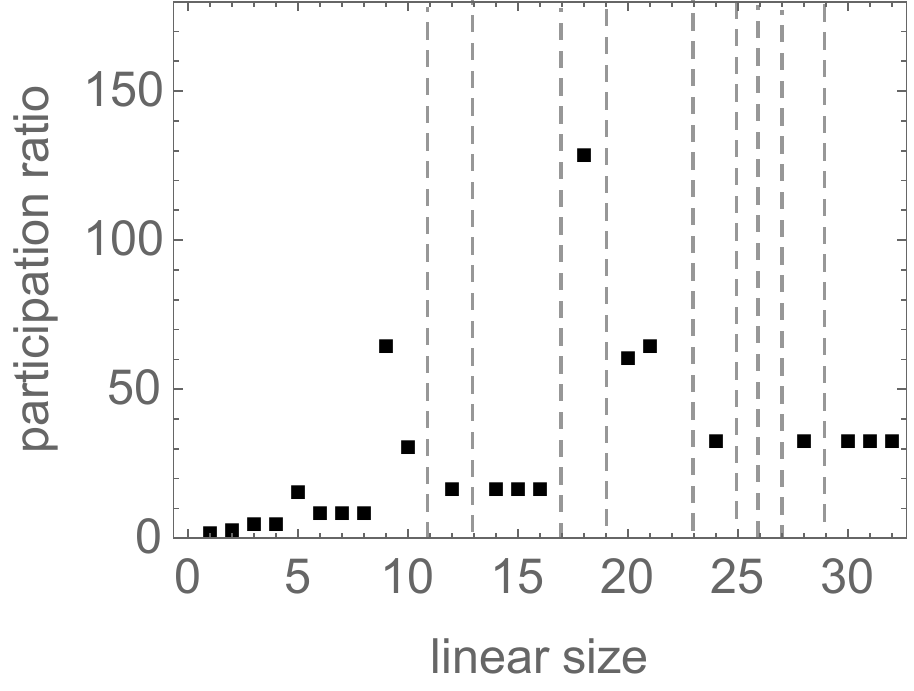}
\end{minipage}
\caption{Properties of the square-lattice parity model. Upper left: Depiction of the first few steps of operator growth for an operator initialized at a site on the A sublattice (dark blue). Subsequent sites are marked in increasingly light shades of blue (for the A sublattice) and red (B sublattice). Upper right: time-averaged Heisenberg operator $X_i$ over 60 timesteps on a $2 \times 100 \times 100$ square lattice (the factor of two is for the two sublattices); the operator (which starts out at the center of the square) spreads but in a structured way. The shading denotes the time-averaged support of the operator on a particular site. Lower left: an example of the various clearly nonrandom structures that arise in the late-time evolution of the $X$ operator. These data are for $2 \times 19 \times 19$ systems, after $80$ timesteps. Lower right: configuration-space IPR vs. linear dimension for this model, showing strong size-dependence.}
\label{squarefig}
\end{center}
\end{figure}

\beq\label{square}
U = \prod_{\langle A B \rangle} \text{CNOT}(A \rightarrow B) \prod_{\langle AB \rangle} \text{CNOT}(B \rightarrow A).
\eeq
i.e., each site on the A sublattice is flipped if an odd number of its (B-sublattice) neighbors are up, but not if an even number are up; then the procedure is repeated, interchanging A and B. 

\subsection{Operator spreading}

Unlike the one-dimensional parity model, this model does \emph{not} appear to be integrable, as all local operators grow under time evolution: an initially compact operator grows at all of its tips, and the geometry of the square lattice does not allow for operators without tips (in other words, there are essentially always sites that touch just one end of the operator). Fig.~\ref{squarefig} shows the first five stages of operator growth in this model. (As we see below in Sec.~\ref{kagome} the Kagome lattice is different in this respect.) As in the previous model, $X$ and $Z$ evolve qualitatively similarly: the dynamics is invariant under flipping $X \leftrightarrow Z$ and interchanging the $A$ and $B$ sublattices. For concreteness we focus on $X(t)$; its growth is shown in Fig.~\ref{squarefig}. A single-site operator spreads, but not uniformly, and there are sets of sites that the operator never reaches. 

\subsection{Eigenstate entanglement and disentangled sites}

Once again, the eigenstates of this model can be explicitly constructed by following cycles of trajectories. The configuration-space IPR shows the same strong and non-monotonic variations that we saw in the Clifford-East model, suggesting that similar commensurability effects are relevant here. Moreover, for the system sizes we have considered, the upper bound from the IPR is generally strong enough to preclude thermal half-chain entanglement in all these systems. 

The single-site entropy is also quite strongly nonthermal in this model: it has a finite entropy of states with disentangled sites, for the same reason as the one-dimensional parity model (Sec.~\ref{1dparity}). If the global configuration is reflection-symmetric about a site, along either axis, the site remains unaffected by the unitary dynamics, and does not entangle with the rest of the system. We emphasize that the present model is not integrable in any obvious sense, and the presence of these disentangled states is a consequence of reflection symmetry. 

\begin{figure}[tbh]
\begin{center}
\begin{minipage}{0.18\textwidth}
\includegraphics[width=0.9\linewidth]{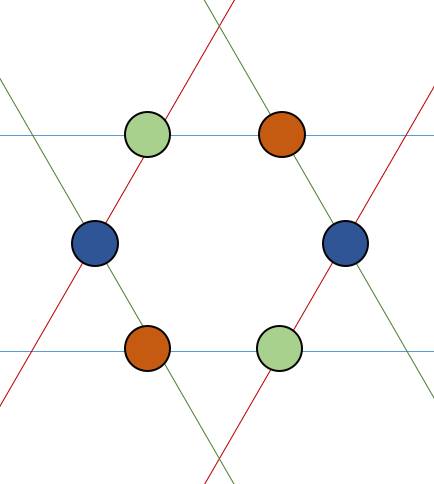}
\end{minipage}
\begin{minipage}{0.25\textwidth}
\includegraphics[width=0.9\linewidth]{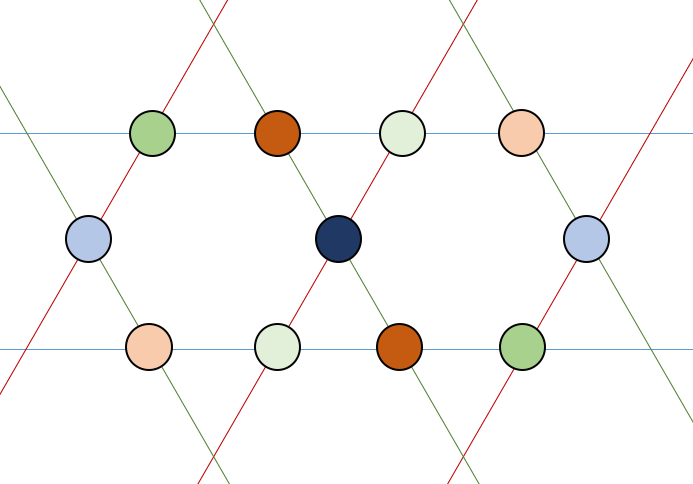}
\end{minipage}
\caption{Left: local conserved operator in the Kagome-lattice parity model; a product of $X$ or $Z$ around a single hexagon is stationary under time evolution. Right: early stages of time evolution of a single $X$ operator, started at the central site in the figure (dark blue). Later stages of evolution are shown in successively lighter colors. After spreading over two hexagons, the operator shrinks back to the origin.}
\label{kagomefig}
\end{center}
\end{figure}

\section{Kagome-lattice parity model}\label{kagome}

The square-lattice parity model does not seem to have local conserved densities; however, to find such conserved densities it suffices to change the lattice geometry to a Kagome lattice. The Kagome lattice has three inequivalent sites, so the corresponding unitary involves three sets of CNOT gates:

\bea
U & = & \text{CNOT}(A, B \rightarrow C) \text{CNOT}(C,A \rightarrow B) \nonumber \\ && \quad \times \text{CNOT}(B,C \rightarrow A).
\eea
There are two types of strictly local conservation laws in this model. First, the product of $X_i$ around a hexagon of the Kagome lattice is invariant under time evolution; second, under time evolution, the product of $Z_i$ around a triangle circulates around one of the hexagons, and returns to itself after two periods. (Note that since the Kagome lattice is not bipartite, the behavior of $X$ and $Z$ is in general different.) 
Thus the Kagome lattice model has extensively many conserved quantities (Fig.~\ref{kagomefig}). Moreover, single-site operators do not spread, but evolve periodically, with period four for the $X$ operator (Fig.~\ref{kagomefig}). The key geometrical difference from the square lattice is that each site adjacent to a hexagon is adjacent to \emph{two} of its vertices, a situation that never occurs in the square lattice. 

\section{Toffoli-gate model}\label{toffsec}

\begin{figure}[tbh]
\begin{center}
\begin{minipage}{0.23\textwidth}
\includegraphics[width=0.9\linewidth]{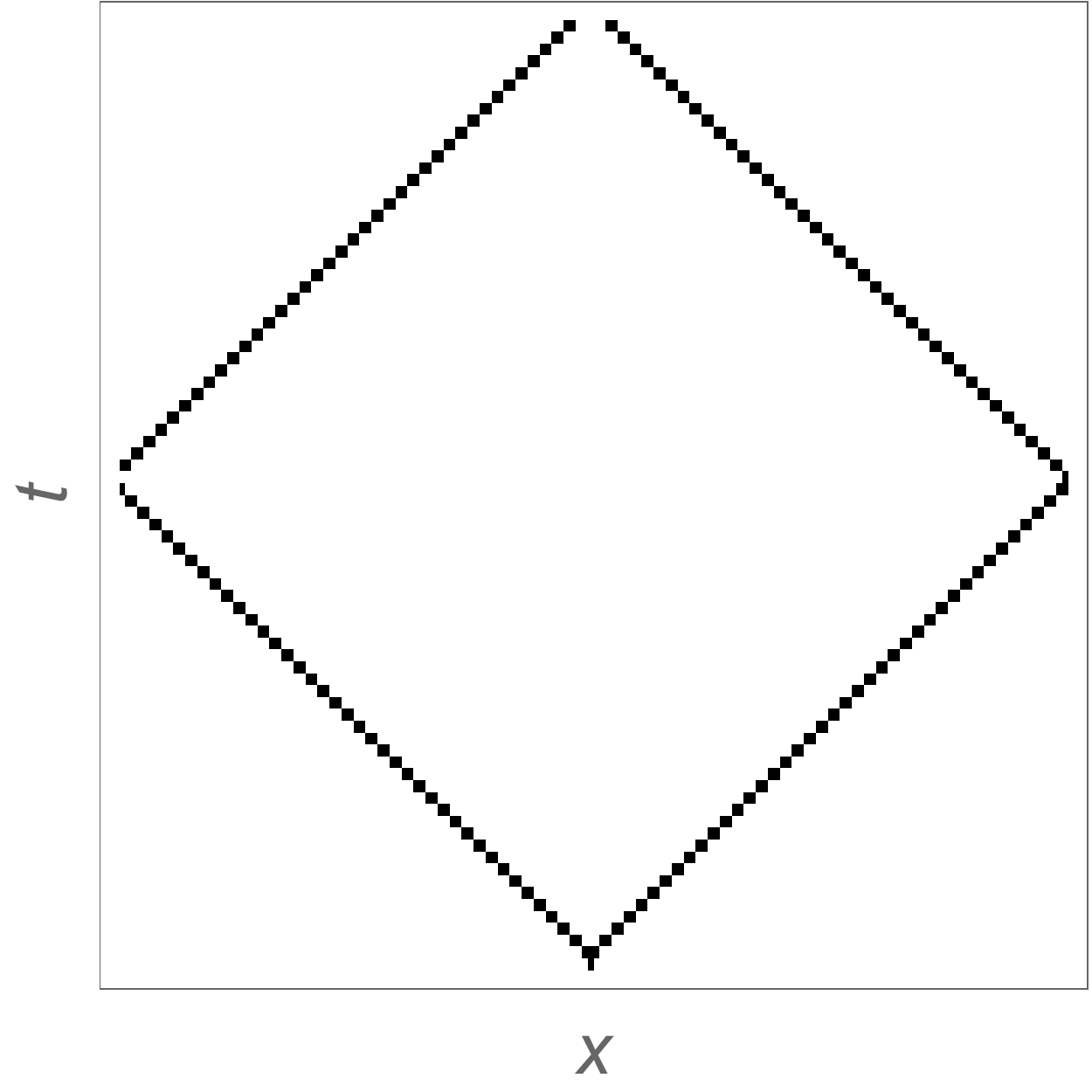}
\end{minipage}
\begin{minipage}{0.23\textwidth}
\includegraphics[width=0.9\linewidth]{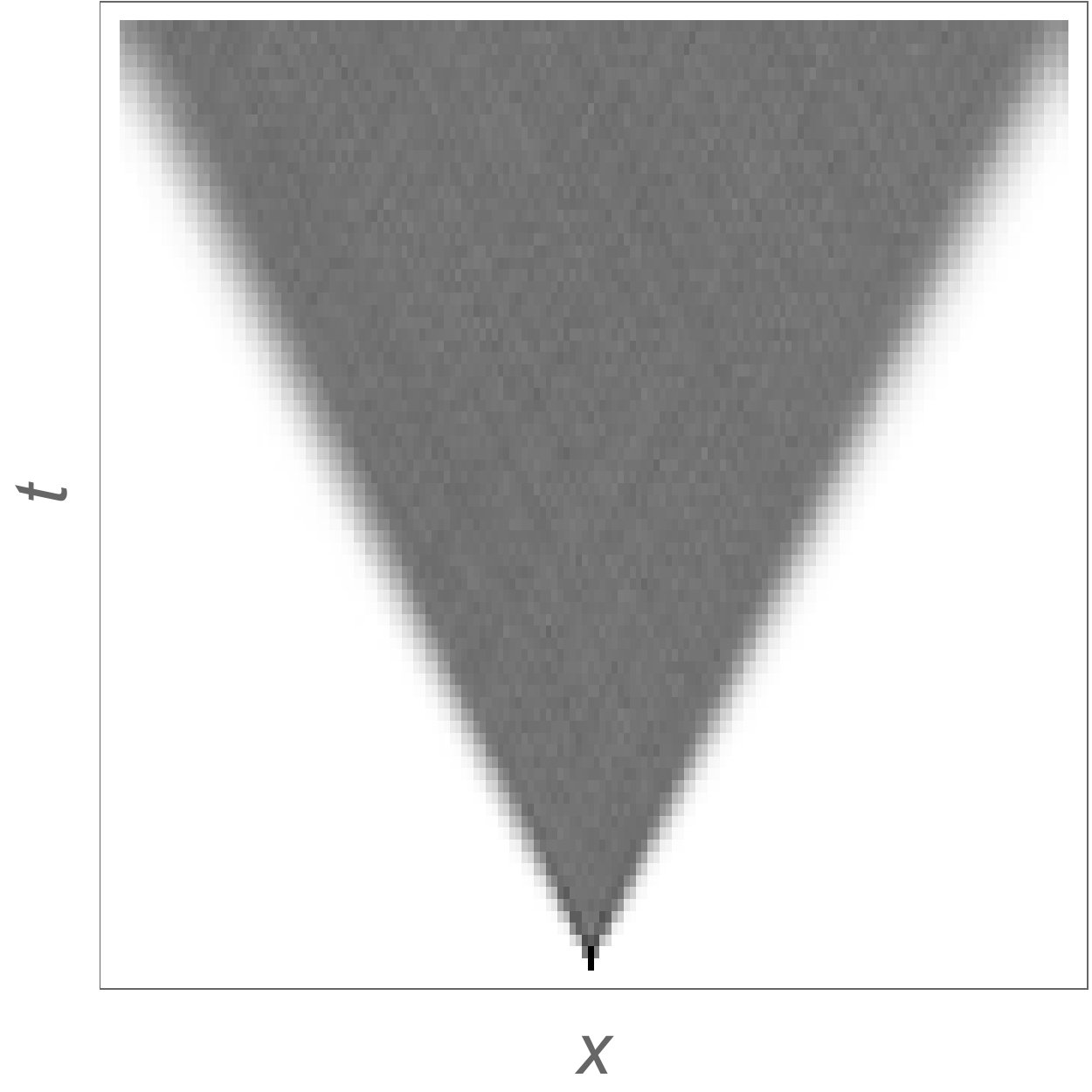}
\end{minipage}
\includegraphics[width = 0.4\textwidth]{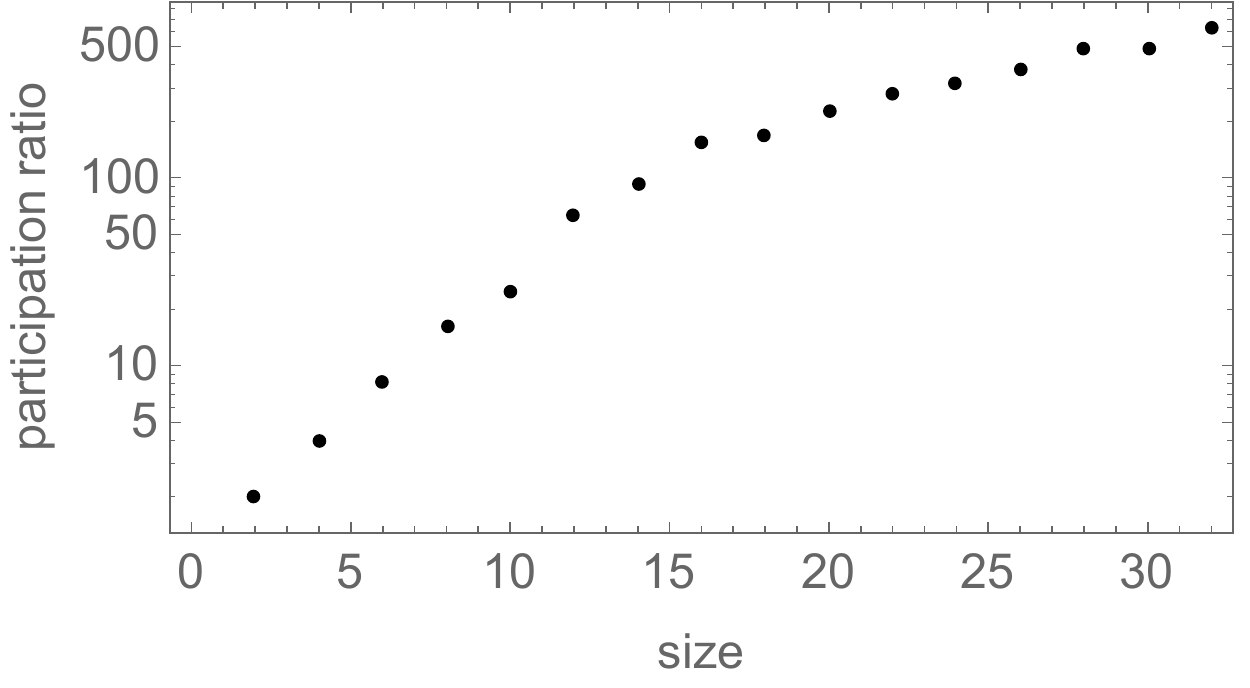}
\caption{Upper panels: Out-of-time-ordered correlator $\langle [Z(t), X(0)]^2 \rangle$ starting from the vacuum state (left) and averaged over $1000$ random initial states (right) in the CNOT-Toffoli model. Darker shading denotes larger OTOC. The averaged OTOC spreads ballistically, but approximately half as fast as a single particle. Lower panel: configuration-space participation ratio~\eqref{ipr} vs. system size for this model.}
\label{tofffig}
\end{center}
\end{figure}

We now briefly turn to a model involving both CNOT and Toffoli gates. A Toffoli gate from bits 1 and 2 to 3, denoted $T(1, 2 \rightarrow 3)$, flips spin 3 if both spins 1 and 2 are up. Using CNOT and Toffoli gates, one can construct a limit of the Frederickson-Andersen model, in which a spin flips if \emph{at least one} of its neighbors is up. The unitary is again made up of two half-cycle unitaries $U(\text{even} \rightarrow \text{odd}) U(\text{odd} \rightarrow \text{even})$, where $U(\text{even} \rightarrow \text{odd}) = T(2,4 \rightarrow 3) \text{CNOT}(2\rightarrow 3) \text{CNOT}(4 \rightarrow 3) T(4,6 \rightarrow 5) \text{CNOT}(4\rightarrow 5) \text{CNOT}6 \rightarrow 5)\ldots$

Toffoli gates do not belong to the Clifford gate set; thus, operator evolution under these gates is not efficiently classically simulable. Nevertheless we can work in the Schr\"odinger picture and follow the evolution of states in the computational basis, which again undergo classical cycles. As before, we can construct eigenstates by tracing classical orbits. 
We find that the half-chain entanglement is subthermal for essentially all system sizes greater than $L = 14$, but grows essentially monotonically without any evident number-theoretic structure. A sharper distinction with the Clifford models has to do with operator spreading. In the Clifford models, even when operators spread, they do so in complex fractal patterns; by contrast, in this model, they fill in the light-cone as they would in a conventional chaotic system, with a state-dependent butterfly velocity (Fig.~\ref{tofffig}).

\section{Exact diagonalization results}\label{edresults}

\begin{figure}[tb]
\begin{center}
\begin{minipage}{0.23\textwidth}
\includegraphics[width = \linewidth]{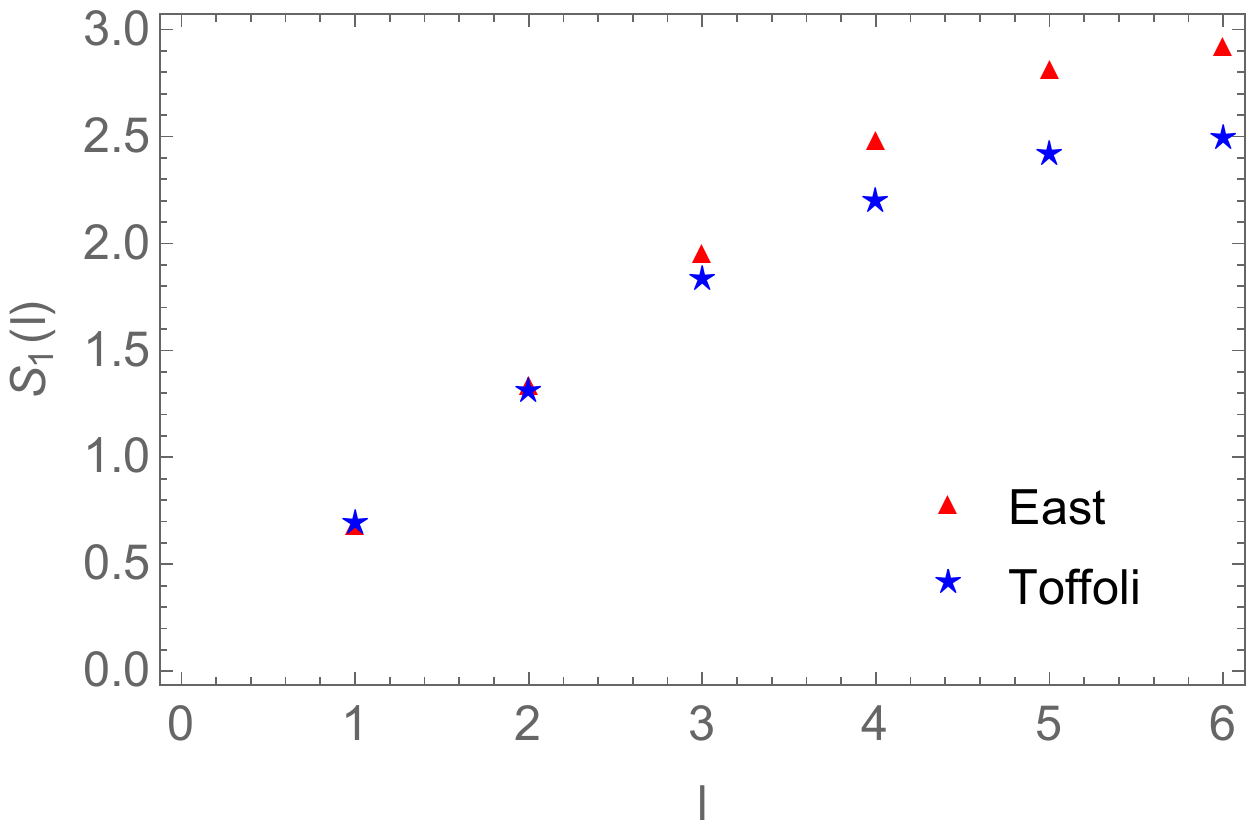}
\end{minipage}
\begin{minipage}{0.23\textwidth}
\includegraphics[width=\linewidth]{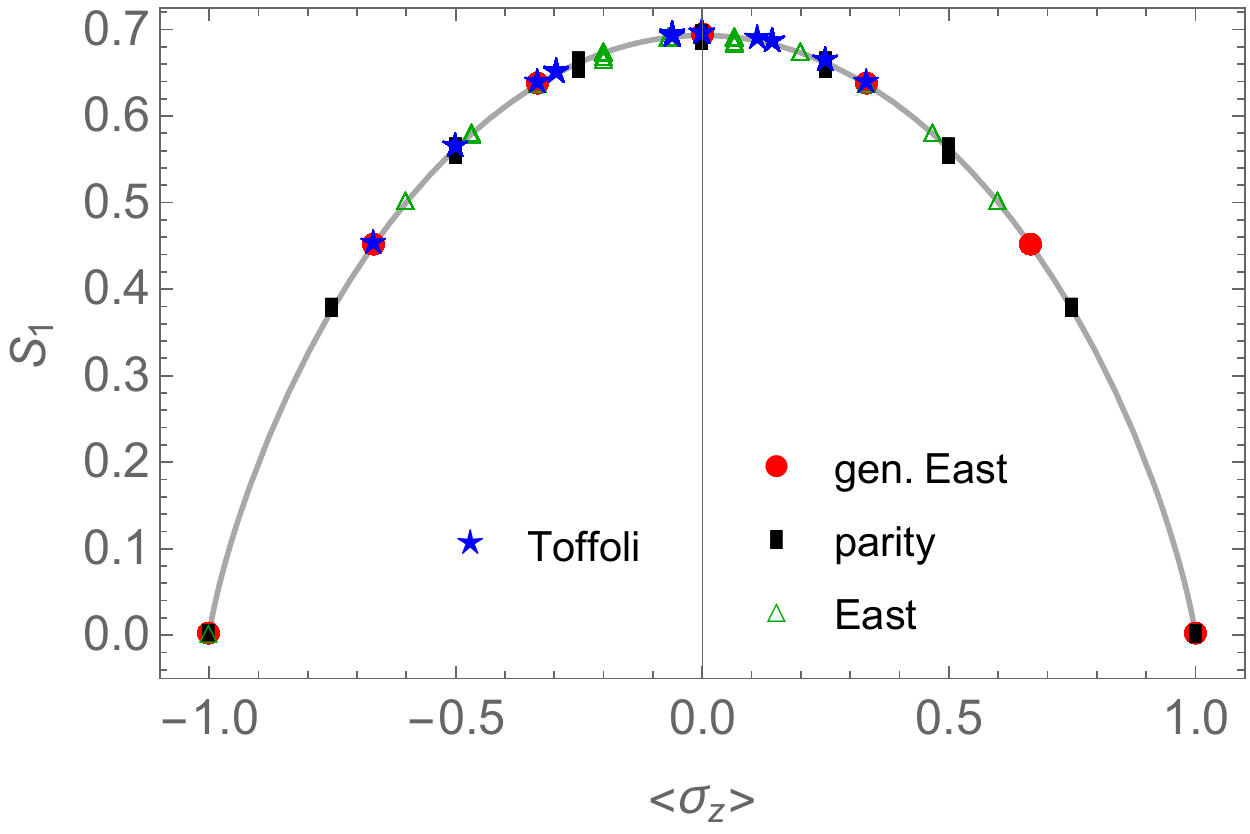}
\end{minipage}
\caption{Exact diagonalization results on $L = 12$ spin chains, for various models. Left: dependence of the entanglement entropy $S_1$ on region size. Right: Scatter-plot of eigenstate expectation values of $\sigma_z$ vs. single-site $S_1$: states have the maximal entanglement consistent with a given $\langle \sigma_z \rangle$. (Note that most of the eigenstates are clustered around the $y$ axis, as canonical typicality would predict.)}
\label{ed}
\end{center}
\end{figure}

We have checked our results against exact diagonalization on small systems up to $L = 12$. The configuration-space IPR~\eqref{ipr} and Renyi entropy match the classical results in all cases we have considered. 
For subsystems much smaller than the half-system, the entanglement appears to grow as a volume law (Fig.~\ref{ed}), then crosses over to slower growth as the system size is increased. An intriguing local property of the eigenstates that emerges from exact diagonalization is that, for a specific eigenstate, the single-site von Neumann entanglement entropy $S_1$ and the expectation value $\langle Z_1 \rangle$ are closely related, in all the models we have considered. Specifically, the entanglement, even in anomalous states, takes the maximum value consistent with $\langle Z_1 \rangle$. 

\section{Discussion}\label{discussion}

We have introduced and analyzed a class of Floquet systems inspired by cellular automata and kinetically constrained models. The dynamics of these systems is essentially classical, and thus tractable, but considered \emph{as quantum spin chains} they have several interesting properties. Two of the models are conventionally integrable, and the other three are not. But even in the nonintegrable models we have considered, operator dynamics and/or eigenstate entanglement differ from both conventional chaotic systems and integrable systems. 
In the nonintegrable Clifford models, operators grow as spacetime fractals. Thus, the behavior of the out-of-time-ordered correlator \emph{inside} the light-cone is anomalous: instead of growing and remaining large, it grows and shrinks in a self-similar pattern. 
Also, one of the models (the Clifford-East model) exhibits one-sided light-cone growth, where the operators $X$ and $Z$ spread respectively to the right and the left. 
By contrast, the model involving Toffoli gates (Sec.~\ref{toffsec}) exhibits fairly conventional light-cone spreading.

More generally, all three of these models seem to exhibit non-thermal eigenstate entanglement for subsystems that are a finite fraction of the full system size. The unusual behavior of the half-chain entanglement entropy stems from the fact that eigenstate entanglement is sensitive to dynamics on timescales that are exponential in system size, and thus inherently has strong boundary effects. These boundary effects are amplified by the classical nature of the dynamics in these models: after $n$ steps, the model only goes through $n$ computational basis states, rather than the generic expectation of $o(2^n)$. Thus in a sense one can think of these models as involving classical motion on the configuration-space hypercube~\cite{BAA}, which takes exponentially many steps to explore the entire hypercube (even though information spreads ballistically in real space). 
It might be that this classicality is what makes ETH fail for extensively large subsystems. We emphasize, however, that there are relatively few theoretical constraints compelling extensively large systems to behave thermally.

One might question whether our results are fully general even for Clifford automata. We believe they are: Ref.~\cite{gutschow2010} classifies Clifford quantum cellular automata, and finds that their operator dynamics is periodic (i.e., localized), has ``gliders'' (i.e., local operators that translate under time evolution, a.k.a. conserved densities), or involves spacetime fractal growth. The models we have studied exemplify all three classes of behavior. From the perspective of thermalization, it is the fractal class that is most interesting. An important question for future work is to determine, first, whether these fractal models are conventional integrable models in disguise or a genuinely different type of model, and second, whether more realistic models exist that exhibit this type of behavior.

\begin{acknowledgments}

We are grateful to Lincoln Carr, David Huse, Vadim Oganesyan, and Andrew Potter for helpful discussions. This work was supported by the NSF Grant No. DMR-1653271. S.G. acknowledges the hospitality of the Aspen Center for Physics, which is supported by NSF grant PHY-1607611.

\end{acknowledgments}


%




\end{document}